\documentclass[aps,amssymb,amsmath,prl,reprint,superscriptaddress,noshowpacs]{revtex4-1}

\usepackage{times}
\usepackage[T1]{fontenc}
\usepackage[latin1]{inputenc}
\usepackage[english]{babel}

\usepackage{graphicx}

\newcommand{\figwidth}{0.45\textwidth} %% specifies size for pictures
\renewcommand\Im{\operatorname{\mathfrak{Im}}}

\newcommand{\affil}{Center for Nanophotonics, FOM Institute for Atomic and Molecular Physics, Science Park 104, 1098 XG Amsterdam, The Netherlands}

\begin{document}
\author{Martin Frimmer}\email{frimmer@amolf.nl}\homepage{http://www.amolf.nl/}
\affiliation{\affil}
\author{Yuntian Chen}
\affiliation{DTU Fotonik, Department of Photonics Engineering, \O rsteds Plads, Building 343, DK-2800 Kongens Lyngby, Denmark}
\author{A. Femius Koenderink}
\affiliation{\affil}
\title{Scanning emitter lifetime imaging microscopy for spontaneous emission control}

\begin{abstract}
We report an experimental technique to map and exploit the local density of optical states of arbitrary planar nano-photonic structures.  The method relies on positioning a spontaneous emitter attached to a scanning probe deterministically and reversibly with respect to its photonic environment while measuring its lifetime. We demonstrate the method by imaging the enhancement of the local density of optical states around metal nanowires. By nano-positioning, the decay rate of a pointlike source of fluorescence can be reversibly and repeatedly changed by a factor of two by coupling it to the guided plasmonic mode of the wire.
\end{abstract}

\date{Submitted: April 11, 2011}
%\pacs{1234567890}
\maketitle %% realizes the layout of the title

Spontaneous emission control is at the heart of photonics, the science of engineering the generation, propagation, and absorption of light.
Since the pioneering work of Purcell  it is known that the emission properties of a spontaneous emitter can be taylored by its optical environment, which determines the number of final states available for the photon emitted in the transition. This is quantified by the local density of optical states (LDOS) \cite{Purcell1946,*Sprik1996}.
Reaching beyond spontaneous emission, the LDOS is a fundamental quantity that also reflects how the electromagnetic  mode structure affects e.g. thermal emission processes, radiation by accelerated charges, and  forces mediated by vacuum fluctuations~\cite{Wilde2006,*Kuttge2009,Novotny2006}.
The LDOS is defined as the imaginary part of the Green's function $\Im\{G\}$~\cite{Novotny2006} and can be thought of as the impedance imposed on a radiating source by its environment~\cite{Greffet2010,Koenderink2010}. It includes all channels offered by the environment, i.e. radiative decay into the far field, decay into confined photonic or polaritonic resonances, and quenching.
Nanophotonic structures exhibiting an LDOS structured at length scales smaller than the wavelength of light include photonic crystals, random scattering materials,
and plasmonic structures, all holding promise to achieve control over all aspects of spontaneous emission, including decay rate   \cite{Lodahl2004,Koenderink2005,Sapienza2010,*Birowosuto2010,Kuehn2006,*Anger2006,*Farahani2005,Akimov2007,*Wei2009,*Fedutik2007,Kolesov2009,Hoogenboom2009}, directionality \cite{Curto2010,Lee2011} and polarization \cite{Moerland2008}.

To unlock the potential of  nanophotonic devices for quantum optics, one requires tools to spatially image the LDOS on a nanometer scale~\cite{Wilde2006,*Kuttge2009,Chicanne2002}. Moreover, to exploit the LDOS to its full potential, it is desirable to first image the LDOS, in order to subsequently position a source deterministically at the optimal location for the actually fabricated structure, as retrieved from the LDOS map.
Drop casting of emitters, randomly or on selectively functionalized substrates, is often used to obtain LDOS data~\cite{Hoogenboom2009,Akimov2007,*Wei2009,* Fedutik2007,Kolesov2009,Curto2010}.
However, in this method emitter positions are fixed after deposition and photonic properties have to be deduced from ensemble averages. Therefore, it is difficult to obtain calibrated LDOS values and impossible to first map the LDOS to then controllably place an emitter in the mapped structure.
Elaborate nano-positioning techniques can assemble photonic devices with single emitters by pushing nano-objects to selected locations~\cite{Schietinger2009, *Huck2011, *Sar2009}. However, such iterative multi-step position-and-probe sequences are time consuming and limit the applicability as an LDOS imaging tool.
These deficiencies can be overcome by scanning probe techniques. For LDOS imaging and simultaneous reversible nano-manipulation of emission, one would prefer a method in which  either light source or photonic structure is attached to the scanning probe.
In pioneering experiments, a single emitter fixed in a substrate was used to image the LDOS of a simple plasmonic structure attached to a scanning probe~\cite{Kuehn2006,*Anger2006,*Farahani2005,Moerland2008}.
Groundbreaking experiments have been performed also in the converse geometry, where the intensity of  emitters attached to sharp probes is monitored while scanning them through the near field of subwavelength structures~\cite{Michaelis2000,*Cuche2010,*Aigouy2003}. Such scanning of pointlike light sources holds great promise for LDOS mapping and nanomechanical manipulation~\cite{Koenderink2005}, since it is directly compatible with the constraints of ubiquitous planar nanofabrication technologies.
%
% %
%
%
In this Letter, we report a nanoscale LDOS imaging technique that combines scanning near field optical microscopy with fluorescence lifetime imaging (FLIM) to map the LDOS around nanoscopic structures by reversible and on-demand positioning of a single nano-sized source of fluorescence. % without requiring a flat surface in the region of interest.
As a paradigmatic example, we investigate Au and Ag metal nanowires, structures of significant interest for plasmon quantum optics \cite{Akimov2007,*Wei2009,*Fedutik2007}. We manipulate the decay rate of a pointlike source of fluorescence reversibly and repeatedly by scanning it relative to a plasmonic nanowire. At selected source positions a significant fraction of decay events is funneled into a guided mode on the wire, proving the possibility to not only control nanomechanically \emph{when} photons are emitted but also \emph{where to}.

\begin{figure}
\includegraphics[width=\figwidth]{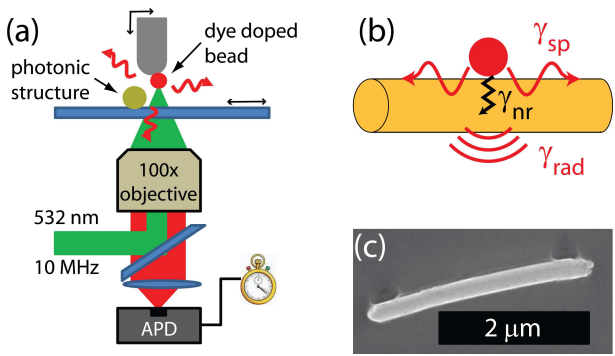}
\caption{(a) Schematic of experimental setup. The fluorescent source is attached to a scanning probe and  positioned with respect to the structure of interest. A pulsed pump laser is focused onto the source whose fluorescence is collected by the same microscope objective. The fluorescence is split off by a dichroic beamsplitter and a color filter and focused onto an APD. (b) Schematic of nanowire with light source in its vicinity. The decay channels, radiative, nonradiative, and into surface plasmons are indicated together with their rates. % $\gamma_{rad},\gamma_{nr},\gamma_{sp}$.
(c) SEM micrograph of a Au nanowire.}
\label{setup}
\end{figure}

Our scanning emitter lifetime imaging microscope is a homebuilt confocal FLIM system based on an inverted microscope, equipped with a scanning probe that addresses the photonic structure from above (Fig.~\ref{setup}a). As a benchmark experiment, we investigate Ag and Au nanowires deposited on a cleaned glass cover slip. The source of spontaneous emission in our experiments, for brevity termed `the source'  in the remainder, is a polystyrene bead with a diameter of 100\,nm, infiltrated with approximately $10^3$ dye molecules with arbitrary orientations, a fluorescence peak at 560\,nm, and a quantum efficiency close to 1 (Invitrogen Fluospheres F8800).
A solution of these beads has been spun on a cleaned  cover slip. The scanning probe, a pulled glass fiber with an end radius of around 100\,nm attached to an xyz piezo arm (piezosystem jena), is held at a constant distance of several nm to the sample surface by shear force feedback \cite{Novotny2006}. This distance and the size of the fluorescing bead minimize the effect of quenching in our experiment~\cite{Anger2006}. We dip the probe into a solution of PMMA in anisole and subsequently approach it to a bead on the sample, which we locate by its fluorescence on a CCD camera. The polymer coating promotes the attachment of a bead to the tip. The light source is pumped by a 532\,nm pulsed laser (Time-Bandwidth), operating at 10\,MHz with a pulse duration $<$10\,ps, focused to a diffraction limited spot on the sample surface by a 100x dry objective (NA 0.95). The fluorescence emitted by the source is collected by the same optics, passes the dichroic beam splitter and an additional long pass filter to be focused onto an avalanche photo diode (APD). The $20\,\mu$m active area of the APD (ID-Quantique id100-20 ULN) and 20x magnification between sample and detector result in a confocal arrangement. The APD is connected to a timing card (Becker \& Hickl DPC-230), recording the arrival times of the laser pulses and the  fluorescent photons.
\begin{figure}
\includegraphics[width=\figwidth]{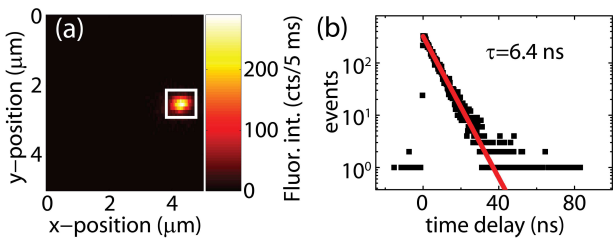}
\caption{(a) Fluorescence intensity map of fluorescing probe scanned across unpatterned cover glass. The bright region corresponds to where the probe passes the laser focus. The signal to background ratio exceeds 10$^3$  at a pump power of 0.2$\,\mu$W.
% Deduce this from an average of 0.2264 cts/px in the first 50 columns of figure a compared to >200 at peak.
(b) Decay trace of photon events from probe positions within box in (a), revealing a single exponential decay with 6.4\,ns time constant (red curve).}
\label{probe}
\end{figure}

\begin{figure*}
\includegraphics[width=\textwidth]{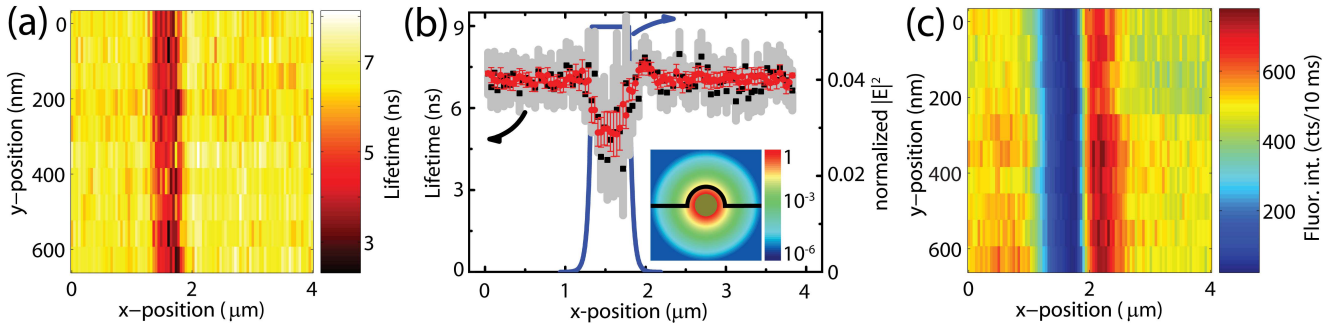}
\caption{(a) Lifetime map of source scanned with respect to Au nanowire. Close to the wire, the lifetime is reduced to half its original value. (b) Black squares: Cross section through first row in (a). Grey errorbars indicate $3\sigma$ confidence interval obtained from the covariance matrix of the fit. Red circles: Lifetime values obtained from averaging all horizontal scan lines in (a) with $3\sigma$ errorbars. Blue line: Intensity of the fundamental wire mode (250 nm diameter)  along the probe trace (55\,nm from wire surface) as indicated in inset~\cite{[{We use finite element modeling, assuming a lossless nanowire (250\,nm diameter, $\epsilon$=-7.911, $\lambda$=560\,nm) in an effective  host index $n$=1.25}. See ] Chen2010}. (c) Fluorescence intensity map obtained during measurement that yielded (a) showing a change in apparent brightness of the source in the vicinity of the wire.}
\label{NWmaps}
\end{figure*}

We  now use the probe to map the LDOS of a photonic structure, a template grown Au nanowire \cite{Lee2006} with a length of several $\mu$m and a diameter of ca. 250\,nm (Fig.~\ref{setup}c).   We position the probe in the laser focus, as established in Fig.~\ref{probe}a, such that the source is continuously excited and its fluorescence detected.
The sample with the wire is now raster scanned underneath the fixed probe. The acquired arrival times of all photons together with the positioning information allow us to determine the decay time of the source for each position relative to the wire. As the main result of this Letter, Fig.~\ref{NWmaps}a shows a spatial map of the lifetime of the source as a function of position with respect to the nanowire. We observe a pronounced reduction in lifetime of the source when it is close to the wire (position  confirmed by simultaneously acquired topography, not shown). The black squares in Fig.~\ref{NWmaps}b show the first row of Fig.~\ref{NWmaps}a, with the grey bars illustrating the $3\sigma$ error interval. While the lifetime of the source is around 7\,ns when it is far from the wire, it drops rapidly to around 4\,ns as soon as the distance between source and wire is of the order of the wire radius. Having passed the wire, the lifetime recovers its original value.
The red points in Fig.~\ref{NWmaps}b take into account all horizontal scan lines in Fig.~\ref{NWmaps}a. This measurement clearly shows that we can reversibly change the excited state lifetime of the source via its position with respect to the nanowire.
The lifetime reported in Fig.~\ref{NWmaps}a is an unambiguous measure for the LDOS, i.e. $\Im\{G\}$, at the emission frequency.
The ability to image $\Im\{G\}$ for any planar nanophotonic system is the main result of this paper. As opposed to position-and-probe techniques~\cite{Schietinger2009}, our method is a real imaging technique with the possibility to repeatedly measure LDOS and calibrate the source in situ.
In contrast to earlier work, where the LDOS around simple plasmon antennas attached to scanning probes was measured by scanning the antenna with respect to an emitter fixed in a substrate~\cite{Anger2006}, our technique can map the LDOS in any planar photonic system, such as ubiquitous lithographically prepared plasmonic and metamaterial systems.
We proceed to interpret the LDOS measured in the particular structure reported here.
 For plasmonic systems, the emitter has three possible decay channels with associated rates, as depicted schematically in Fig.~\ref{setup}b. These are decay into a photon, $\gamma_{rad}$, into a plasmon, $\gamma_{sp}$, and direct nonradiative decay due to Ohmic losses in the metal, $\gamma_{nr}$. The lifetime measured in Fig.~\ref{NWmaps}a is the inverse of the total decay rate $\gamma=\gamma_{rad}+\gamma_{nr}+\gamma_{sp}$, which is proportional to $\Im\{G\}$, i.e. the LDOS. The doubling of the decay rate is a clear indication of the increased LDOS in the vicinity of the nanowire. The spatial extent of the lifetime reduction is of the order of the wire radius, as was theoretically predicted \cite{Chang2007} on the basis that the change in decay rate occurs mostly due to coupling of the emitter to a guided plasmonic mode of the nanowire.
This scaling is confirmed by the blue line in Fig.~\ref{NWmaps}b, showing the calculated normalized intensity of the fundamental mode~\cite{Chen2010}.
The magnitude of the measured LDOS enhancement is comparable to reported values for single NV-centers in diamond nanocrystals of size $\approx$50\,nm attached to Ag nanowires \cite{Kolesov2009}. One might have expected that the number of molecules distributed over the entire volume of the fluorescing bead is a disadvantage for mapping the LDOS compared to nanodiamonds containing single emitters. Our results show that regarding positioning accuracy for mapping the LDOS, a 50\,nm diamond nanoparticle, even with only a single NV center, is not necessarily superior to a dye doped bead of 100\,nm.
In our view,  the relevant \emph{optical} dimension of the source is the product of refractive index and particle size, which is comparable for 100\,nm polystyrene and 50\,nm diamond particles.

Figure \ref{NWmaps}c shows the integrated number of detected photons as a function of probe position from the same scan that yielded Fig.~\ref{NWmaps}a. Similar fluorescence intensity maps were obtained in earlier experiments with fluorescent scanning probes~\cite{Michaelis2000}.
%The data set yielding Fig.~\ref{NWmaps}a can also be plotted as a fluoresence intensity map (Fig.~\ref{NWmaps}c), similar to those obtained in earlier experiments with fluorescent scanning probes \cite{Michaelis2000}.
The wire in close proximity to the source suppresses its apparent brightness by up to ten times, while there is a region of enhanced fluorescence on the right hand side of the wire. The spatial width of these features is of the order of the wire diameter.
Such fluorescence intensity data is much more complicated to interpret than lifetime maps as Fig.~\ref{NWmaps}a \cite{Michaelis2000,Taminiau2008b}, since it is a convolution of pump field, collection efficiency and rate enhancements.
The wire  causes an enhancement of pump field (polarized perpendicular to the wire axis) on the wire sides and a suppression behind the wire, similar to the case of metallic Mie-spheres~\cite{Koenderink2010}. We attribute the fluorescence enhancement in Fig.~\ref{NWmaps}c to such pump field enhancement. The asymmetry can be explained by an asymmetry in the attachment of the bead to the scanning probe. It is not a result of scan direction, since consecutive rows in Fig.~\ref{NWmaps}c are acquired with alternating directions.
Besides pump field suppression, non-radiative channels offered by the wire reduce the observed fluorescence intensity.
We argue that the dominant non-radiative decay process in our system is the generation of plasmons, since quenching occurs at emitter-metal distances of only few nanometers \cite{Anger2006}, much smaller than our source diameter and source-wire separation.
Therefore, both the observed change in decay rate and in fluorescence intensity are likely not due to quenching, but rather indicate that a significant fraction of decay events is into plasmons.
\begin{figure}
\includegraphics[width=\figwidth]{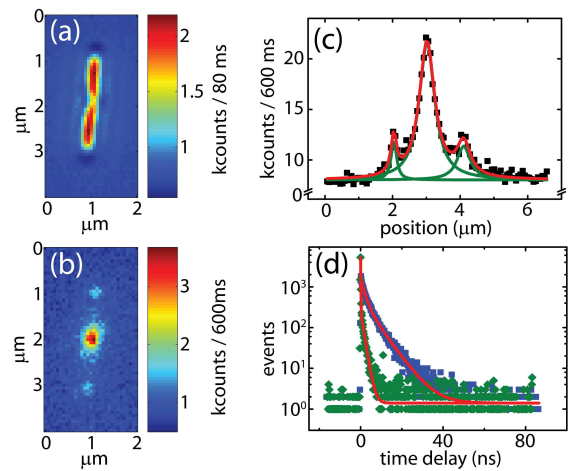}
\caption{(a) Ag wire with attached bead in white light illumination imaged on CCD camera. (b) Fluorescence image of structure in (a) under epi-illumination by pump laser. Light emerges from the  bead and from wire ends. (c) Black points: Cross section along wire obtained by binning the central 10 columns in (b). Red line: Fit to data with three peaks (green lines). (d) Decay traces of light source on probe away (blue squares) and after deposition on Ag nanowire (green diamonds). Red lines are biexponential fits.}
\label{AgWire}
\end{figure}

A quantitative measure for the fraction of decay events into plasmons compared to decay into free space is most easily derived from a complementary experiment, in which we used our scanning probe to deposit a fluorescing bead on a Ag nanowire with a length of about $2\,\mu$m and a diameter of 300\,nm (BlueNanoInc, SLV-NW-300). Single-crystalline Ag nanowires are superior to Au for this purpose due to their longer plasmon propagation lengths, while the plasmonic mode structure of Ag and Au wires is comparable~\cite{Chen2010}. The deposited bead shows up as a faint signature from the scattered intensity in the wire center on a CCD camera under white light  illumination (Fig.~\ref{AgWire}a).
In the fluorescence image from the same system under laser epi-illumination the fluorescing bead appears as a bright source of emission, while the wire ends are also bright (Fig.~\ref{AgWire}b). This confirms that the emitters indeed decay into a plasmon that can only couple to free space at irregularities such as the wire ends.
A cross section along the wire shows the central peak from the photons emitted into free space and two smaller ones from the wire ends, corresponding to quanta emitted into a plasmon (Fig.~\ref{AgWire}c). The data (black points in Fig.~\ref{AgWire}c) are fitted well (red line) with three peaks (green lines).
From the areas under the peaks we conclude that the $\beta$ factor, i.e. the fraction of quanta emitted into the wire mode, is around 20\% where we have neglected plasmon loss en route to the wire ends.
We also measured the decay  of the source on the probe away from the wire (blue squares in Fig.~\ref{AgWire}d) and after coupling to the wire (green diamonds).
Evidently, the lifetime of the source is drastically reduced by positioning it on the wire. We have fitted both curves with a biexponential decay (red lines in Fig.~\ref{AgWire}d). The source on the probe is fitted excellently with a slow component of 6.2\,ns, a relative amplitude weight of 28\%, and a fast component of 1.8\,ns. The same source on the wire has a slow component of 1.4\,ns, a weight of 4.8\%, and a fast decay of 0.1\,ns.
A conservative estimate for the rate enhancement is 4.4 (ratio of slow components). Another estimator is the first moment of the time traces (4.3\,ns off vs. 0.5\,ns on the wire), which yields an enhancement ratio of 9.
Both values are higher than the one obtained in the scanning emitter experiments which can be attributed to the fact that there the source is kept at a distance of several nm above the surface.

In conclusion, we presented a technique to map $\Im\{G\}$ of arbitrary photonic structures with nanometer resolution.  Such scanning emitter lifetime imaging is suited to exploit the backaction of the photonic environment on a spontaneous emitter for a plethora of structures of current interest, hitherto inaccessible to established near and far field techniques.  Our work shows repeated switching of the decay rate of a pointlike light source by a factor of two by reversible and on demand positioning of an emitter within its nanoscale photonic environment. This constitutes a major step towards full nanomechanical control over all aspects of spontaneous emission, including decay rate, directionality and spectral composition. In the limit of a single scanning  quantum system, our  method will even give access  to the reverse process of spontaneous emission, i.e. the absorption of single photons in the vicinity of nanostructures \cite{Celebrano2011,*Gaiduk2010}, as well as position dependent coupling, energy transfer and photon-photon correlations between emitters linked by a nanophotonic structure \cite{Martin-Cano2010}.
\begin{acknowledgments}
We thank H. Schoenmaker, M. Seynen, and I. Attema for technical support, A. Polman for supplying the Au wires, D. van Oosten, M. Burresi, and L. Kuipers for fruitful discussions.
This work is part of the research program of the ``Stichting voor Fundamenteel
Onderzoek der Materie (FOM),'' which is financially supported by the
``Nederlandse Organisatie voor Wetenschappelijk Onderzoek (NWO).''
\end{acknowledgments}

\bibliography{FrimmerManuscript}

\end{document}